\input{aipcheck}
\documentclass[final]{aipproc}

\layoutstyle{8x11single}

\def\l{{\cal L}}
\def\d{\partial}
\def\m{{\cal M}}
\def\sz{{S^1}/{Z_2}}
\def\ot{\otimes}
\def\sus{SU(6)}
\def\sut{SU(3)}

\def\uo{U(1)}
\def\sutc{SU(3)_C}
\def\sutl{SU(2)_L}
\def\uoy{U(1)_Y}
\def\suth{SU(3)_H}
\def\uoc{U(1)_C}
\def\uob{U(1)_B}

\begin{document}

\title{Dynamical symmetry breaking of SU(6) GUT in $5-$dimensional spacetime with orbifold $\sz$}


\author{A. Hartanto}{
  address={Indonesian Center for Theoretical and Mathematical Physics, Jl. Ganesha 10, bandung 40132, Indonesia}
  ,altaddress={Theoretical Physics Laboratory, THEPI Research Division\footnote{http://www.fi.itb.ac.id}, Bandung Institute of Technology, Jl. Ganesha 10, bandung 40132, Indonesia}
}
\author{F. P. Zen}{
  address={Indonesian Center for Theoretical and Mathematical Physics, Jl. Ganesha 10, bandung 40132, Indonesia}
  ,altaddress={Theoretical Physics Laboratory, THEPI Research Division\footnote{http://www.fi.itb.ac.id}, Bandung Institute of Technology, Jl. Ganesha 10, bandung 40132, Indonesia}
}
\author{J. S. Kosasih}{
  address={Indonesian Center for Theoretical and Mathematical Physics, Jl. Ganesha 10, bandung 40132, Indonesia}
  ,altaddress={Theoretical Physics Laboratory, THEPI Research Division\footnote{http://www.fi.itb.ac.id}, Bandung Institute of Technology, Jl. Ganesha 10, bandung 40132, Indonesia}
}
\author{L.T. Handoko\thanks{laksana.tri.handoko@lipi.go.id, handoko@teori.fisika.lipi.go.id}}{
  address={Group for Theoretical and Computational Physics, Research Center for Physics\footnote{http://teori.fisika.lipi.go.id}, Indonesian Institute of Sciences, Kompleks Puspiptek Serpong, Tangerang 15310, Indonesia}
  ,altaddress={Department of Physics\footnote{http://www.fisika.ui.ac.id}, University of Indonesia, Kampus UI Depok, Depok 16424, Indonesia}
}

\begin{abstract}
The symmetry breaking of $5-$dimensional $\sus$ GUT into $4-$dimensional $\sut \ot \sut \ot \uo$ with orbifold $\sz$ through Scherk-Schwarz mechanism is investigated. It is shown that the origin of Little Higgs can be generated to further break $\sut \ot \sut \ot \uo$ down to the electroweak scale through Higgs mechanism.
\end{abstract}

\maketitle

\section{Introduction}
\label{intro}

The unification of electroweak theory and strong interaction within a universal framework based on gauge theory with large symmetry remains an open problem after decades. Although the electroweak $\sutl\ot \uoy$ and the strong $\sutc$ theories are in impressive agreements with most of experimental observables \cite{pdb}, both are independently constructed. The model, known as the Standard Model (SM) with symmetry $\sutc \ot \sutl\ot \uoy$, is lacking of explaining the unification of three gauge couplings at a particular scale assuming  that our nature should be explained by a single unified theory, the so called grand unified theory (GUT).

In the last decades, some efforts have been dedicated to investigate  gauge theories with larger gauge symmetries inspired by the successfull electroweak theory. Those theories assumed the gauge invariance under particular symmetries larger than the SM's one, but contain $\sutc \ot \sutl \ot \uoy$ as a part of its subgroups at electroweak scale. One of them is the GUT model based on $\sus$ group \cite{hartanto}. The model argues that the electroweak scale physics is realized through breaking patterns $\sus \rightarrow \sutc \ot \suth \ot \uoc$, and subsequently $\suth \rightarrow \sutl \ot \uob$. Here, $H$ denotes a new quantum number called as hyper-isospin, while  the combination of quantum numbers induced by $\uob$ and $\uoc$ reproduces the familiar hypercharge associated with $\uoy$ in the electroweak theory. 

In spite of its potential to be a new alternative of non-supersymmetric GUT beyond the well-known $SU(5)$, unfortunately the model is suffered from the difficulty to find an appropriate Higgs multiplet to realize the above mentioned breaking patterns through Higgs mechanism \cite{hartanto2, hartanto3}. It can be concluded that the allowed Higgs multiplets in the model are not able to reproduce all particle spectrums within the experimental bounds.

Following recent progress on extra dimension physics, non-Higgs mechanisms to break the symmetry have been pointed out in some previous works. Some of them proposed the so-called Scherk-Schwarz mechanism to dynamically break the symmetry induced by the orbifold of extra dimension \cite{nomura1, nomura2, quiros}. Rather than directly breaking the symmetry, the effect on compactifying the extra dimension is considered to induce the Higgs bosons itself. This approach is known as the Higgs -- gauge boson unification \cite{nomura3}.

In this paper, both mechanisms are utilized to resolve the problem of breaking the $\sus$ GUT. The first breaking pattern is realized by compactification of extra dimension in 5-dimensional $\sus$, and at the same time it generates the Higgs bosons to induce the second one. Moreover, the arising Higgs bosons are chosen to be the so-called little Higgs with $T-$parity symmetry \cite{schmaltz}. This scenario is to forbid tree-level contributions from new physics to the electroweak observables, but still allowing the loop effects to cancel the divergences and raising the cut-off of electroweak theory up to 10 TeV \cite{cheng}.

The paper is organized as follows. First the symmetry breaking to $\sut \ot \sut \ot \uo$ due to dimensional reduction with orbifold $\sz$ is explained in the next section. Subsequently it is shown that the arising gauge bosons from broken $5-$dimensional $\sus$ could be considered as the little Higgs to induce the second breaking. Finally a short summary and discussion on the remaining problems are given.

\section{Symmetry breaking in 5-dimensional spacetime $\m^4 \ot \sz$}
\label{sec:5su6}

There are two known mechanisms for breaking the symmetry in a non simply-connected spaces as $\m^4 \ot (\sz)$ where compactification is performed on the orbifold, that is Scherk-Schwarz anf Hosotani mechanisms. This paper deploys the first one. 

The Scherk-Schwarz mechanism is based on twist $T_i$ that represents a discrete group $G$ defining the orbifold by means of local or global symmetry of the lagrangian under consideration. For $5-$dimensional case compactified on $\sz$ orbifold, the Scherk-Schwarz twist is $T_i = \mathrm{exp}(\pi i \omega t^i)$ with twisted field \cite{quiros},
\begin{equation}
 \phi(x,y+2\pi R) = \mathrm{e}^{2\pi i \omega^i t^i} \, \phi(x,y) \; ,
  \label{eq:twist}
\end{equation}
where $t^i$ is the generator with a given direction in generator space, while $\omega^i$ is the corresponding parameter. In order to break the $\sus$ symmetry by $5-$dimensional kinetic term, the following conditions,
\begin{eqnarray}
 \{ \omega^{\hat{a}} t^{\hat{a}}, Z_2 \} = \omega^{\hat{a}} \{ t^{\hat{a}}, Z_2 \} & = & 0 \; ,  \\
	\left[ t^a, Z_2 \right] & = & 0 \; ,
 \label{eq:condition}
\end{eqnarray}
must be satisfied with $t^{\hat{a}}$ denotes the off-diagonal generators ($\hat{a} = 9, \cdots, 26$). These lead to three solutions : 1) $\omega^{\hat{a}} = 0$ and $\{ t^{\hat{a}}, Z_2 \} \neq 0$; 2) $\omega^{\hat{a}} \neq 0$ and $\{ t^{\hat{a}}, Z_2 \} = 0$; 3) $\omega^{\hat{a}} = 0$ and $\{ t^{\hat{a}}, Z_2 \} = 0$. The last solution is more prefered in the current case, since the second one leads to non-periodic fields, while the first one obviously keeps the symmetry.

Applying the above last solution to $\sus$ yields $\{ t^{\hat{a}}, Z_2 \} = 0$, 
\begin{equation}
	Z_2 = \left(
	\begin{array}{cccccc}
	1 & 0 &  & &  & \\
	0 & 1 & &&& \\
	& & 1 & 0 & & \\
	& & 0 & -1 & & \\
	& & & & -1 & 0 \\
	& & & & 0 & -1 \\
	\end{array}
	\right) \; .
	\label{eq:z}
\end{equation}
This matrix actually represents the boundary conditions of orbifold $\sz$ to break $5-$dimensional $\sus$ into $4-$dimensional $\sut \ot \sut \ot \uo$. The unbroken parts are determined by $[t^a, Z_2] = 0$ ($a = 1, \cdots, 8, 27, \cdots, 35$).

In $5-$dimensional spacetime, the lagrangian for gauge sector is written as, 
\begin{equation}
	\l_A = -\frac{1}{2} \, F_{MN} F^{MN} \; ,
\end{equation}
where $F_{MN} = F_{MN}^i t^i$, $F_{MN}^i \equiv \d_M A_N^i - \d_N A_M^i + g \, f^{ijk} A_M^j A_N^k$ and the covariant derivative is $D_M \equiv \d_M - i g \, A_M^i t^i$. Using Lie algebra $[ t^i, t^j] = i \, f^{ijk} t^k$ and keeping the  lagrangian invariant, $t^i$ can be naturally splitted into unbroken parts $t^a$ and broken parts $t^{\hat{a}}$. These results lead to parities of gauge bosons $A_M^i$ as written in Tab. \ref{tab:parity}. Here, $A_\mu^a$'s are gauge bosons on $4-$dimensional brane, while $A_y^{\hat{a}}$'s are massless scalar bosons. Both have zero-modes and are non-vanishing. Therefore the orbifold breaking split parities into even or odd. In the current case, the unbroken $4-$dimensional gauge bosons $A_\mu^a$ and the broken extra-dimensional gauge bosons $A_y^{\hat{a}}$ are even functions with the first zero modes are $4-$dimensional vector particles, while the last zero modes are $4-$dimensional massless scalars. The unbroken extra-dimensional gauge bosons $A^a_y$ are odd functions with no zero mode representing $5-$dimensional massless scalars. On the other hand, the broken $4-$dimensional gauge bosons $A_\mu^{\hat{a}}$ obtain the masses from VEV, $\left\langle \Phi_0 \right\rangle = \left\langle A^{\hat{a}}_{y,0} \right\rangle = \omega/R$.

One finds out that NGB's are basically $A_y^{\hat{a}}$ which can be expressed in $5-$dimensional even scalars as \cite{nomura2},
\begin{equation}
 A_y^{\hat{a}} = \tilde{\Phi}_+(x,y)
	= \frac{1}{\sqrt{\pi R}} \Phi_+^{(0)}(x) 
	+ \frac{1}{\sqrt{\pi R}} \sum_{n=1}^{\infty} \Phi_+^{(n)}(x) \, \cos \left( \frac{n y}{R} \right) \; ,
\end{equation}
and in $5-$dimensional odd scalars as,
\begin{equation}
 A_y^a = \tilde{\Phi}_-(x,y)
	= \frac{1}{\sqrt{\pi R}} \sum_{n=1}^{\infty} \Phi_-^{(n)}(x) \, \sin \left( \frac{n y}{R} \right) \; .
\end{equation}
It should be noted that $A_\mu^a$ belongs to $4-$dimensional $\sut \ot \sut \ot \uo$ symmetry, while $A_\mu^{\hat{a}}$ becomes massive.

\begin{table}[t]
 \begin{tabular}{p{3cm}p{2cm}p{2cm}}
  \hline
  parity & \multicolumn{2}{c}{gauge boson} \\
  \hline
  even & \multicolumn{1}{c}{$A_\mu^a$} & \multicolumn{1}{c}{$A_y^{\hat{a}}$} \\ 
  odd & \multicolumn{1}{c}{$A_\mu^{\hat{a}}$} & \multicolumn{1}{c}{$A_y^a$} \\ 
  \hline
 \end{tabular}
 \caption{Parities of gauge bosons $A_M^i$.}
 \label{tab:parity}
\end{table}

\section{The candidate of Little Higgs in $4-$dimensional spacetime}
\label{sec:higgs}

Now let us consider a type of little Higgs model, namely the simplest little Higgs \cite{schmaltz}. The lagrangian for Higgs sector in $5-$dimensional $\m^4 \ot (\sz)$ is,
\begin{equation}
 \l_\Phi = ( D^M \Phi^\dagger ) (D_M \Phi) \; ,
  \label{eq:lh}
\end{equation}
with $\Phi$ is a sextet of scalar bosons with the elements $\Phi_j$ ($j = 1, \cdots, 6$).

As already discussed in the preceeding section, the symmetry is broken by Scherk-Schawrz mechanism and the boundary conditions on orbifold $\sz$ are given by Eq. (\ref{eq:z}). Furthermore, the breaking is under conditions : $\omega \neq 0$ or $\omega = 0$, and $\{ t^{\hat{a}}, Z_2 \} = 0$. Putting an additional condition for $\tilde{\Phi}(x,y)$,
\begin{equation}
 \tilde{\Phi}(x, y + 2 \pi R) = \tilde{\Phi}(x,y) \; ,
\end{equation}
and using Eq. (\ref{eq:twist}) yield,
\begin{equation}
 \Phi(x,y) = \mathrm{e}^{{i \omega^{\hat{a}} t^{\hat{a}} y}/R} \tilde{\Phi}(x,y) \; .
  \label{eq:transf}
\end{equation}

Transforming the lagrangian in Eq. (\ref{eq:lh}) with \ref{eq:transf} gives,
\begin{equation}
 \l_{\tilde{\Phi}} = \left( D^\mu \tilde{\Phi}^\dagger \right) 
           \left( D_\mu \tilde{\Phi} \right) 
         + \left( D^y \tilde{\Phi}^\dagger \right) 
           \left( D_y \tilde{\Phi} \right) 
         + \frac{{\omega^{\hat{a}}}^2}{R^2} \mathrm{Tr} \left( {t^{\hat{a}}}^2\right) \tilde{\Phi}^\dagger \tilde{\Phi}
         + i \frac{\omega^{\hat{a}}}{R} \left[ \tilde{\Phi} \left( D^y \tilde{\Phi}^\dagger \right) - \tilde{\Phi}^\dagger \left( D_y \tilde{\Phi} \right) \right]
         - V\left(\tilde{\Phi}^\dagger\tilde{\Phi}\right) \; .
\end{equation}
Obviously, the kinetic term is not invariant, while the potential remains the same since $\Phi^\dagger \Phi = \tilde{\Phi}^\dagger\tilde{\Phi}$. Recalling Eq. (\ref{eq:z}) and anticommuting relation $\{ t^{\hat{a}}, Z_2\} = 0$, again the unbroken generators are $t^a$ and satisfy $[ t^a, Z_2] = 0$. This proves that the remaining conserved symmetry at this stage is $\sut \ot \sut \ot \uo$.

The breaking of $\sus$ to $\sut \ot \sut \ot \uo$ yield 18 broken generators  $t^{\hat{a}}$ where each of them corresponds to one massless NGB. Furthermore, requiring the potential $V$ to be invariant under $\sus$ leads to,
\begin{equation}
 \sum_{j,k} \frac{\d V}{\d \tilde{\Phi}_j} \, t^i_{jk} \tilde{\Phi}_k = 0 \; .
\end{equation}
At the vacuum $\tilde{\Phi}_0$, taking the differential with respect to $\tilde{\Phi}_l$ one obtains,
\begin{equation}
 \sum_{j,k} m_{lj}^2 \, t^i_{jk} (\tilde{\Phi}_k)_0 = 0 \; .
\end{equation}
Hence, for $t^i_{jk} (\tilde{\Phi}_k)_0 \neq 0$ one gets $m_{lj}^2 = 0$ as massless NGB bosons. In other words, the relation $t^i_{jk} (\tilde{\Phi}_k)_0 \neq 0$ constraints the number of NGBs. 

Without loss of generality, one can choose two VEVs correspond to twisted fields (non-periodic, broken) $\tilde{\Phi}_3$ and $\tilde{\Phi}_6$ as follows,
\begin{equation}
 \left( \tilde{\Phi}_3 \right)_0 = \left(
 \begin{array}{c}
  0 \\
  0 \\
  \left\langle 0 \left| \tilde{\Phi}_3 \right| 0 \right\rangle \\
  0 \\
  0 \\
  0
 \end{array}
 \right) = \left(
 \begin{array}{c}
  0 \\
  0 \\
  {\omega^\prime}/{R} \\
  0 \\
  0 \\
  0
 \end{array}
 \right) \equiv v^\prime 
  \; \; \; \; \; \; \; , 
  \; \; \; \; \; \; \; 
 \left( \tilde{\Phi}_6 \right)_0 = \left(
 \begin{array}{c}
  0 \\
  0  \\
  0 \\
  0 \\
  0\\
  \left\langle 0 \left| \tilde{\Phi}_6 \right| 0 \right\rangle
 \end{array}
 \right) = \left(
 \begin{array}{c}
  0 \\
  0  \\
  0 \\
  0 \\
  0\\
  {\omega^{\prime\prime}}/{R}
 \end{array}
 \right) \equiv v^{\prime\prime}   \; .
\end{equation}
Since $t^i \, v^\prime \neq 0$ and $t^i \, v^{\prime\prime} \neq 0$, there should be 22 NGBs. These can be used to construct the $\sus$ little Higgs, for example,
\begin{equation}
 \frac{1}{f} \left(
 \begin{array}{cc}
  (0)_{3 \times 3} & \left(
    \begin{array}{cc} 
     (0)_{2 \times 2} & (h)_{2 \times 1} \\
     ({h^\prime}^\dagger)_{1 \times 2} & 0 
    \end{array}
    \right) \\
 \left(
    \begin{array}{cc} 
     (0)_{2 \times 2} & (h^\prime)_{2 \times 1} \\
     (h^\dagger)_{1 \times 2} & 0 
    \end{array}
    \right) & (0)_{3 \times 3} 
 \end{array}
 \right) \; ,
\end{equation}
where $h$ and $h^\prime$ are the would-be SM Higgs doublet, while $f$ is the root-square of quartic sum of VEVs.

\section{Summary}
\label{sec:summary}

A scenario to break $5-$dimensional $\sus$ GUT down into $4-$dimensional $\sut \ot \sut \ot \uo$ symmetry through Scherk-Schwarz mechanism is explained. At the same time it also generates the candidate for Higgs bosons which break spontaneously the subsequent symmetry into electroweak scale.

Further works on deriving the physical gauge bosons at electroweak scale and analysing some experimental constraints are in progress and will be published elsewhere.

\begin{theacknowledgments}
The work is partially supported by the Riset Kompetitif LIPI in fiscal year $2009$. AH thanks the Indonesian Center for Theoretical and Mathematical Physics, and Group for Theoretical and Computational Physics, Research Center for Physics, Indonesian Institute of Sciences for warm hospitality during this work.
\end{theacknowledgments}

\bibliographystyle{aipproc}
\bibliography{icanse20092}

\end{document}